\documentclass[12pt]{article}
\usepackage{amsmath,amsfonts,amssymb,amsthm}
\usepackage{mathtext}
\usepackage{epsfig,enumerate}

\newtheorem{thm}{Theorem}

\newtheorem{lem}{Lemma}

\newcommand{\RR}{\mathbb R}

\begin{document}

\centerline{\bf P.Grinevich, S.Novikov \footnote{P.Grinevich, Landau Institute for Theor Physics, pgg@landau.ac.ru; S.Novikov, Landau Institiute
 for Theor Physics and IPST/Math, University of Maryland, College Park, novikov@ipst.umd.edu}}

\vspace{0.5cm}
{\bf Discrete $SL_2$ Connections and Self-Adjoint Difference Operators}
 {\bf on the Triangulated 2-manifolds }

\vspace{1cm}

{\bf Abstract}. {\it Discretization Program of the famous Completely Integrable Systems and associated Linear Operators
was developed in 1990s. In particular, specific properties of the second order difference operators on the 
triangulated manifolds and equilateral triangle lattices
were studied in the works of S.Novikov and I.Dynnikov since 1996. They involve factorization
of operators, the so-called Laplace Transformations, new discretization of Complex Analysis and
 new discretization of $GL_n$ connections on the triangulated $n$-manifolds.
The general theory of the new type discrete $GL_n$ connections was developed. 
However, the special case of $SL_n$-connections (and unimodular  $SL_n^{\pm}$ connections such that $\det A=\pm 1$ )
was not selected properly. As we prove in this work, it  plays fundamental role (similar to magnetic field 
in the continuous case)  in the theory of  self-adjoint discrete Schrodinger operators for the equilateral 
triangle lattice in $\RR^2$. In Appendix~1 we present a complete characterization of rank 1 unimodular $SL_n^{\pm}$ 
connections. Therefore we correct a mistake made in the previous versions of our paper (we wrongly claimed that 
for $n>2$ every unimodular $SL_n^{\pm}$ Connection is equivalent to the standard Canonical Connection).  
Using communications of Korepanov we completely clarify connection of classical theory of electric chains 
and star-triangle with discrete Laplace transformation on the triangle lattices\footnote {The works of S.Novikov 
with and without collaborators quoted here can be found in his homepage $www.mi.ras.ru/\tilde{}snovikov$ click
Publications, items 136,137,138,140,146,148,159,163, 173,174,175.}.

\vspace{0.3cm}

{\bf 1. Discrete $GL_n$ Connections.}

\vspace{0.2cm}

Let $M^n$ be any simplicial complex in particular triangulated manifold equipped by the set of (real positive) numbers $u_{T:P}\neq 0$
assigned to every $n$-simplex $T$ and its vertex $P\in T$. They define an operator $Q$ mapping functions of vertices to the functions of $n$-simplices
$$Q\psi_T=\sum_{P}u_{T:P}\psi_P$$

{\bf Definition 1}. The Discrete $R^+$-valued $GL_n$ Connection is defined by the equation $Q\psi=0$ with $u_{T:P}>0$ in $R$. Its coefficients are defined up to the arbitrary Abelian Gauge Transformations determined by the pair of positive nonzero functions of triangles and vertices $g_T> 0,h_P> 0$:
$$\psi\rightarrow h_P\psi_P,\mu^T_{PP'}\rightarrow h_{P'}/h_P\mu^T_{PP'}, Q\rightarrow g_TQ/h_P$$

{\bf Remark.} The case of $GL_2$ connections in the 2-dimensional simplicial skeletons of 3-manifolds may present special interest.

{\bf Remark.} We will develop a theory of $C^*$-valued connection in the next work.

The elementary {\bf parity connection} is defined as a mapping assigning to every closed thick path
$\gamma$ its ``parity''
$P(\gamma)=(-1)^{m(\gamma)}$, where $m(\gamma)$ is the number of the simplices in the thick path
$\gamma$. As we know, the parity connection is trivial if and only if the $n$-simplices can be colored
in black and white. In previous works we called such colorings {\bf discrete conformal structures.}
In all cases below we speak without further comments about $SL_n$ connections if parity connection
is trivial, and we speak about $SL_n^{\pm}$ connections if parity connection is nontrivial (sometimes
the sign $\pm$).

The {\bf Canonical Connection} is defined by the requirement that all coefficients $u_{T:P}$ are equal to $1$.

  Only ratios $$\mu_{PP'}^T=u_{T:P}/u_{T:P'}>0$$ are important for the connection $Q\psi=0$. The so called {\bf ''framed path''} is defined as
 $\gamma^{fr}=[P_0P_1...P_m; T_1,T_2...,T_m]$ where $P_{i-1}P_i$ is an edge in the $n$-simplex $T_i, i=1,...,m$.

  For every connection framed paths have {\bf an Abelian Framed Holonomy Representation}  assigning to every framed path a number $$\gamma^{fr}\rightarrow \prod_i \mu^{T_i}_{P_{i-1}P_i}=\mu(\gamma^{fr})$$
 The semigroup $\Omega_{framed}(M,P_0)$ of the closed framed paths with initial vertex $P_0$ maps into the multiplicative group $k^*$ of our basic field.  We consider here  $R^+\subset k^*=R^*$ only:$$\mu: \Omega_{framed}\rightarrow k^*$$
  The initial point is unimportant because everything is abelian. {\bf The expressions $\mu(\gamma^{fr})$ are gauge invariant.} The simplest of them
  are $$\rho_{PP'}^{TT'}=\mu_{PP'}^T\mu_{P'P}^{T'}=\rho_{P'P}^{T'T}=\mu^{T}_{PP'}/\mu^{T'}_{PP'}$$ corresponding to the framed paths $\gamma^{fr}=[PP'P;TT']$.

   {\bf The Reconstruction of Connection for $n=2$}. In order to reconstruct connection we need to know all $\rho_{PP'}^{TT'}$ and also collection of $\mu(\gamma_j^{fr})$ representing the set of generators of the group $H_1(M,Z)$. Corresponding procedure is presented in the work \cite{N4}. For $n=2$ this procedure is especially simple, especially for the oriented manifolds. We describe it below.

   Let $M$ be an oriented 2-manifold.

  Define quantity  $$\rho(T)=\prod_{R_q\in T}\rho^{TT_q}_{R_q}$$ where $T_q\bigcap T=R_q=ij, T=ijk$, $R_q$ is an edge in $T$.
  This multiplicative 2-cochain is homologous to zero as it was proved in \cite{N4}. For the noncompact case it is obvious.
  For the compact closed orientable manifold we  proved that
  $$\prod_{T\in M}\rho(T)=1$$
  This result easily follows from definition of $\rho(T)=\prod_{R_i}\rho^{TT_{i}}_{R_i}$ where $R_i$ is an oriented edge in $T$ and $T_i\bigcap T=R_i$. We have $\rho^{TT'}_{ij}\rho^{TT'}_{ji}=1$ which
  implies that our product is equal to one, and multiplicative cochain is cohomologous to zero.

  Let $\delta \lambda=\rho^{-1/2}>0$, so we have $$\lambda_{ij}\lambda_{jk}\lambda_{ki}=\rho^{-1/2}(T), T=ijk,\lambda_{ij}\lambda_{ji}=1$$

  Finally we put
  $$\mu^T_{ij}=\lambda_{ij}\cdot(\rho_{ij}^{TT_q})^{1/2}$$

  This formula solves our problem
  following the work \cite{N4}.

The cochain $\lambda$ is nonunique: one can multiply it by the arbitrary co-cycle. However, they can be changed in the cohomology classes because  the abelian  gauge transformations
of the connection   acts exactly in that way. It leads to the  invariants of connections in the  group $H^1(M)$ additional to the ''local'' quantities like $\rho^{TT'}_{ij}$.

  {\bf A Nonabelian Holonomy Representation} is defined by the closed ''Thick Paths''. By definition, a thick path $\gamma^{thick}=T_1...T_m=T_0$ is realized by the sequence of $n$-simplices where $F_i=T_i\bigcap T_{i+1}$ is a common $n-1$-face,  $F_i\neq F_{i+1}$. Every closed thick path defines a linear map
  $$K:\Omega_{thick}(M,T_0)\rightarrow GL_n(R),$$
solving the equation $Q\psi=0$ in every simplex $T_i$ where data are given at the face $F_i$ by induction, starting from $T_0=T_m$.

  {\bf A Nonabelian Curvature } in the point $P$ for  $n=2$ is defined as a Nonabelian Holonomy corresponding to the Thick Path $\gamma_P^{thick}=T_1...T_m$ consisting of 2-simplices $T_i=PP_{i}P_{i+1}$ surrounding  vertex $P$--see Fig 1

   \vspace{0.3cm}

\begin{figure}
\begin{center}
\mbox{\epsfxsize=5cm \epsffile{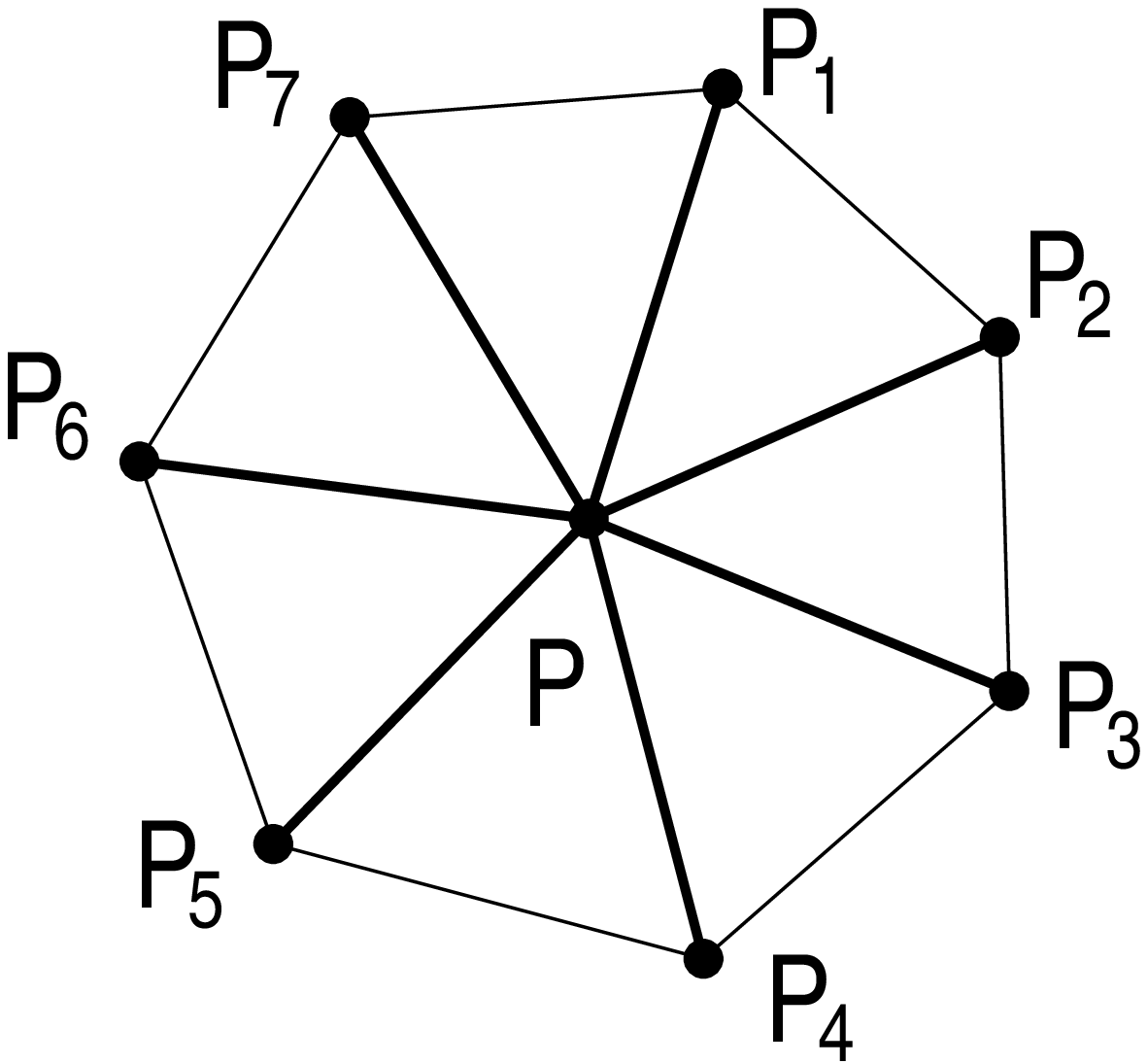}}
\end{center}
\label{fig:fig1}
\caption{}
\end{figure}

 The star of this vertex is $St(P)=T_1\bigcup T_1\bigcup ...\bigcup T_m$.
  Nonabelian Curvature is an upper triangle $2\times 2$-matrix $K(P,T_1)$ with diagonal  $(1,\pm\mu_P)$ where $\mu_P=\mu(\gamma_P^{fr}))$ and second row $(\alpha(P_i,P),\pm\mu_P$). Here $\gamma_P^{thick}$
  is a closed thick path $\partial St_P$ starting and ending in any vertex $P_i\in T_i\in St_P, P_i\neq P$. Easy to see that following lemma is true:

  \begin{lem}For the closed paths surrounding vertex $P$  we have
  $$\mu_P=\prod_{i=1,...m}\rho_{PP_i}^{T_{i}T_{i+1}}$$ where index $i$ is varying modulo $m$.
  \end{lem}
  In the work \cite{N4} one can find all expressions for the coefficients $\alpha(P_i,P)$ of the Nonabelian Curvature through the gauge invariant quantities $\mu^T_{PP'}\rho^{TT'}_{PP'}$.

{\bf Remark}. For all $n>2$ we define nonabelian curvature in the
 same way using the star $St(\sigma^{n-2})$ of every $n-2$-simplex. We use a thick path consisting of all $n$-simplices of this star.
 It leads to the upper triangle matrix with 2 nontrivial elements only, exactly as in the case $n=2$.

  {\bf Problem:} How to construct effectively all discrete  $SL_2^{\pm}$ connections on the triangulated manifold $M^2$?

  {\bf  Construction.} We use the following procedure. Assign to every edge $R \in M$ a number $A(R)>0$.  For every triangle $T\in M$
  we have exactly 3 vertices $P_1,P_2,P_3$ and 3 edges $R_1^T,R_2^T,R_3^T$. Here $R_i=P_{i}P_{i+1}$, and $i$ is counted modulo $3$.
  We define the Connection Operator $Q$ by the equations $$Q\psi_T=\sum_iu_{T:P_i}\psi_{P_i}$$  $$ u_{T:P_i}u_{T:P_{i+1}}=A(P_{i}P_{i+1})$$
  \begin{thm}1.Discrete Connection defined by such operator $Q$ is locally and globally $SL_2^{\pm}$.

  2.Every  $ SL_2^{\pm}$ connection  can be obtained
  by that construction.
  \end{thm}

{\bf Proof of the first part}. For every edge $R=PP'$ and 2 triangles $T=PP'S,T'=P'PS'$ where $T\bigcap T'=R$ we defined above a gauge invariant function $$\rho^{TT'}_{PP'}=
u_{T:P'}^{-1}u_{T:P}u_{T':P'}u_{T':P}^{-1}$$ Taking into account  relations between the products of the connection coefficients,
we are coming to the following result
$$\rho^{TT'}_{PP'}=u_{T:P}^2/u_{T':P}^2=u_{T':P'}^2/u_{T:P'}^2$$
In particular it means that functions $\rho$ for the edges $PP'$ can be calculated from data in one vertex only.
 Consider the series of triangles $T_1,T_2...,T_k$ with the same vertex $P\in T_i=PP_{i}P_{+1}i$ forming a thick path
$T_i\bigcap T_{i+1}=PP_i$. We have
$$\prod_{i=l}^{i=k}\rho^{T_{i-1}T_i}_{PP_{i-1}}=\mu (\gamma^{fr})$$ where $$\gamma^{fr}=[PP_l...P_{k-1}P_kP;T_{l-1}T_{l-1}T_l...T_{k-1}T_kT_k]$$

\vspace{0.3cm}

\begin{figure}
\begin{center}
\mbox{\epsfxsize=3cm \epsffile{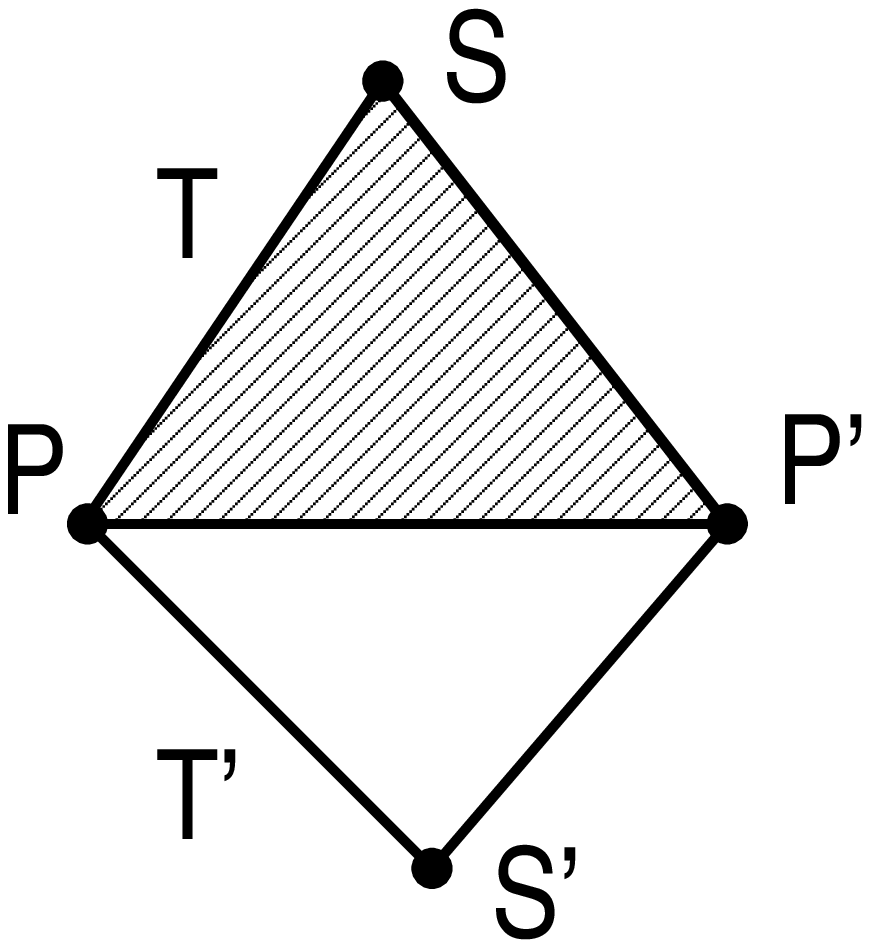}}
\end{center}
\label{fig:fig2}
\caption{}
\end{figure}

\vspace{0.3cm}

For the whole cycle $\partial St(P)$ around the vertex $P$ this equality implies  $\mu(\gamma^{fr})=1$, because $l=1, P_1=P_k$ in this case. The edge $PP_1$
disappear from the path. So our connection is locally $SL_2^{\pm}$.

In order to prove  that the connection is globally $SL_2^{\pm}$, we consider any thick path $\gamma^{thick}=T_0T_1...T_m$ starting from the edge $R_0=P_0P'_0$.
Following the work \cite{N4}, every thick path is defined by the word in the free associative semigroup with 2 generators $a_1,a_2$: either this word starts from $a_1$ and looks like
$a_1^{p_1}a_2^{q_1}....$ or it starts from $a_2$, i.e we have $a_2^{q_1}a_1^{p_1}...$ where  $q_j,p_j$ are positive integers, $j=1,2,...k$.
We present this picture as a pair of 2 vertical boundary half-lines starting from the
initial vertices $P_0$ and $P'_0$ of the form $P_0P_1P_2...$ and $P'_0P'_1P'_2...$. This half-lines $l,l'$ with the initial edge  form the boundary of the
''diagram'' defining   thick path (i.e. the triangulated strip in the half-plane above the edge $P_0P'_0$) where all edges are either ''vertical''--i.e.
belong to the half-lines $l'l'$
or ''horizontal''--i.e. join $l$ and $l'$--see Fig 3.

\begin{figure}
\begin{center}
\mbox{\epsfxsize=6cm \epsffile{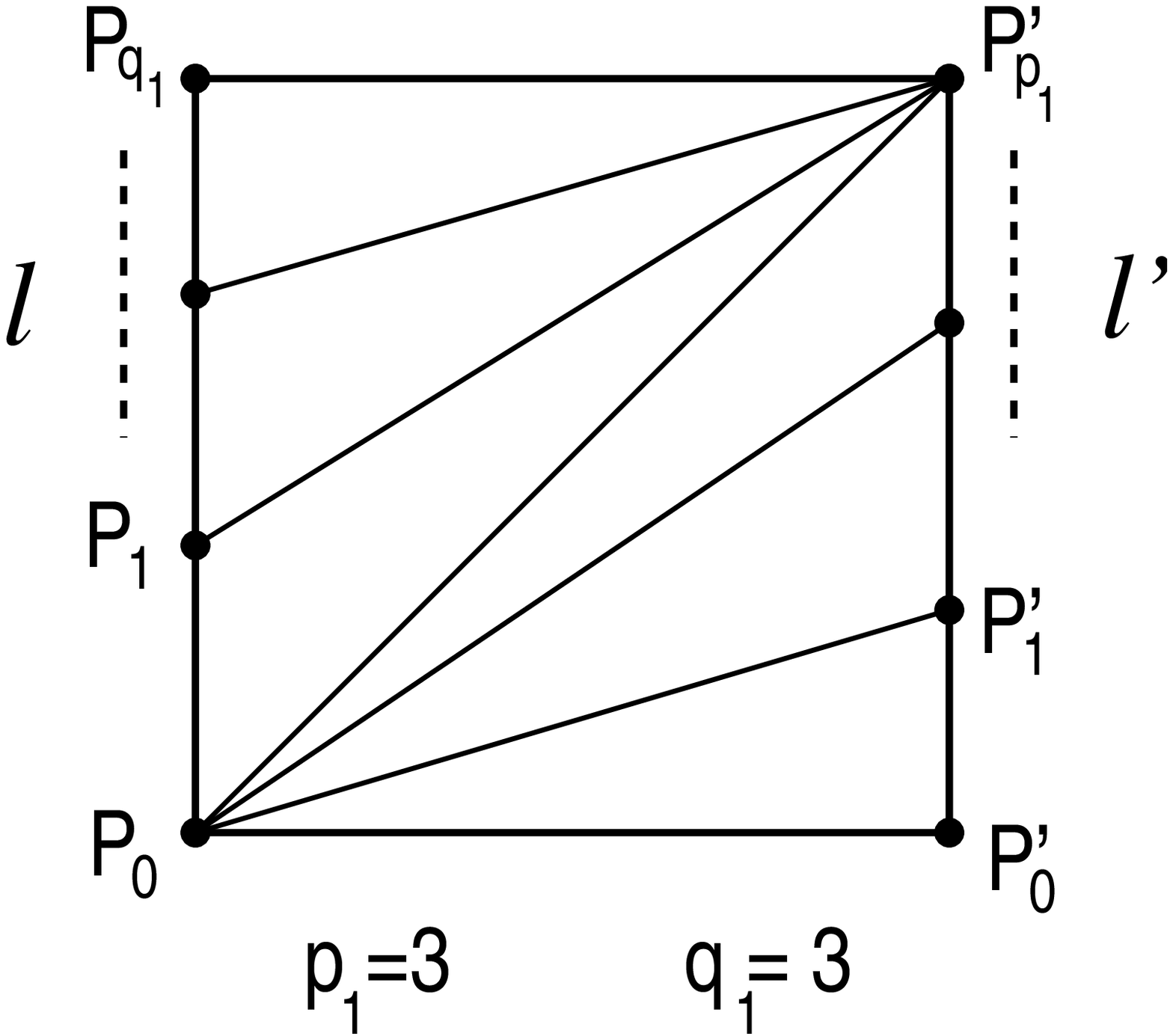}}
\end{center}
\label{fig:fig3}
\caption{}
\end{figure}

\vspace{0.3cm}

We are going to apply the  ''product equality'' for the horizontal $\rho$:
$$\prod_{R_j}\rho_{R_j}^{T_{j-1}T_j}=\mu(\gamma^{fr})$$
where $R_j$ are all horizontal edges in the closed thick path and $\mu(\gamma^{fr})=\mu(l)\mu(l')$. Here $l',l$ are the boundary framed closed paths for the thick path with framing
given by the triangles $T_j$ in the strip.

 As we see, at first we are going around the first center $P_0\in l$ exactly $p_1$
times, after that we shift center to the  the vertex $P'_{p_1}$ at the line $l'$ and rotate around it $q_1$ times, and so on. For each such step we use the previous product formula.
The quantities $\rho^{TT'}_{PP'}$ in this case can be expressed as a ratios of the  squares of coefficients of the connection operator $Q$ attached to the same vertex $P$ or $P'$ from the right and left sides of this edge. So we can jump from the line $l$ to the line $l'$ and back in process of calculation, when
corresponding rotation  ends.
Finally, for the closed thick path we are coming to the same initial coefficients but in the opposite powers like in the case of curvature. So the product is equal to one. Our theorem is proved.

{\bf Proof of the second part for simply connected manifolds}:

 Let some  $SL_2^{\pm}$ connection on simply connected manifold be given
as a collection of connection coefficients $\mu^T_{PP'}>0$. We choose an arbitrary edge $R=PP_1$ and number $A(PP_1)>0$.
 As a first step we define coefficients $A(PP_i)$ for the edges  $PP_i$ in the star $St(P)$. They are uniquely defined by the connection coefficients $\mu^{T}_{PP'}=u_{T:P}/u_{T:P'}$. To calculate them we  use the assumption that the products of  coefficients $$u_{T':P}u_{T':P_i}=u_{T'':P}u_{T'':P_i}, T''\bigcap T'=PP_i$$   of the connection operator $Q$ along the edge $PP_i$ for the triangles $T',T''$ from the right and left sides of the edge $PP_i$ are equal to each other. We are solving the equation for $u$ through $\mu$ step by step for $i=1,2,...$.
  using the first coefficient $A(PP_1)$,  going around $P$. It can lead to contradiction only after passing the full circle $P_1...P_m$ around $P$. But  the connection is $SL_2^{\pm}$. So the individual simplicial star does not lead to contradiction.  After that we extend it adding triangles by the following rules:\\
  1.Add new triangle which  has only 2 common vertices and one common edge with the set of previous triangles.\\
  2.Add new triangle which has exactly 3 common vertices with the set of previous triangles and 2 common edges.

  In the first case we obviously have no source for contradiction extending connection operator $Q$.
  In the second case our new triangle is the last triangle in the simplicial star of the vertex $P$. No contradiction appears here because our connection is $SL_2^{\pm}$. So we meet
   no contradictions at all, and  we are staying
in the simply connected class of domains.  We  find finally all coefficients of the operator $Q$.
Our Theorem is proved for simply connected manifolds.

{\bf Proof of the second part for the non-simply-connected manifolds}:

Consider now non-simply-connected manifold
$M$ and its universal covering $p:N\rightarrow M$. We have $\pi_1(N)=1$. Our connection in $M$ canonically defines a connection in $N$ with same coefficients $\mu^{T'}_{i'j'}=\mu^T_{ij}$ where $T'\rightarrow T$ is a canonical linear isomorphism provided by covering map $p(T')=T$. Choose initial edge $E_1=PP'\subset N$ and coefficient $A(PP')$ as above.
Construct the whole field of coefficients $A(E)$ for all edges $E$ in $N$ as above, because $N$ is simply connected.
We have a regular covering group action of $g\in \pi_1(M)$ in $N$ where $g:N\rightarrow N$. Define all images
$E_g=g(E_1)\subset N$. Let $A(E_g)=\lambda_g^2, \lambda_g>0$. Now we construct representation acting on the coefficients of the operator $Q$. This representation preserves connection coefficients $\mu^T_{ij}$ which are ratios of the operator coefficients in every triangle $T$. So we have
$g:\pi_1(M)\rightarrow R^+$ where $g\rightarrow \lambda_g$.
This representation gives us   determinants of the holonomy maps realized by the thick paths. For all $SL_2^{\pm}$-
connections ($R^+$-valued) in $M$  we have $\lambda_g=1$ by definition. Therefore we obtain a field  $A(E)$, and coefficients of operator $Q$ are well-defined for $M$.
Our Theorem is proved.

{\bf Remark}. {\it Every locally $SL_2^{\pm}$ connection can be considered as $SL_2^{\pm}$ connection in universal covering $N$
with operators $Q$ given by the set of edge numbers $A(E)$ and coefficients $u_{T:P}$ such that for $g\in \pi_1(M)$ we have
$$u_{gT:gP}=\lambda_g u_{T:P}$$ or $g^*Q=\lambda_g Q$
for some representation $\pi_1(M)\rightarrow R^+$.}

\vspace{0.3cm}

{\bf 2. $SL_2$-Connections and Difference Self-Adjoint Operators.}

\vspace{0.2cm}

 Consider here any triangulated 2-manifold $M^2$ with ''discrete conformal structure'' which is defined as a
coloring of 2-simplices (triangles) into the black and white colors, according to  the works \cite{ND3,N5,N7,GNRoma}. As we observed in \cite{N1,ND1,N2}, every scalar real self-adjoint second order operator $L$ can be presented here in the factorized form (''black'' and ''white''):

$$L\psi_P=\sum_{P'}b_{P:P'}\psi_{P'}=\sum_{P\neq P'}b_{P:P'}\psi_{P'}+W(P)\psi_P$$ where either $P=P'$ or  $PP'$ is an edge.
 Let $b_{P:P'}=b_{P':P}$. We call the set of coefficients $b_{P:P}$
''Potential'' $W(P)=b_{P:P}$ and require $b_{P:P'}> 0$ for all  $PP'$. There exist unique operators $Q^b$ and $Q^w$--the black and white triangle operators--such that $$L=Q^{b+}Q^{b}+W^b=Q^{w+} Q^w+W^w$$

Here
$Q^b\psi_T=\sum_{P\in T}u_{T:P}\psi_P$ where $T$ is any black triangle, and $Q^w_T=\sum_{P\in T}u_{T:P}\psi_P$ where $T$ is any white triangle (see Fig 4).

\vspace{0.2cm}

\begin{figure}
\begin{center}
\mbox{\epsfxsize=6cm \epsffile{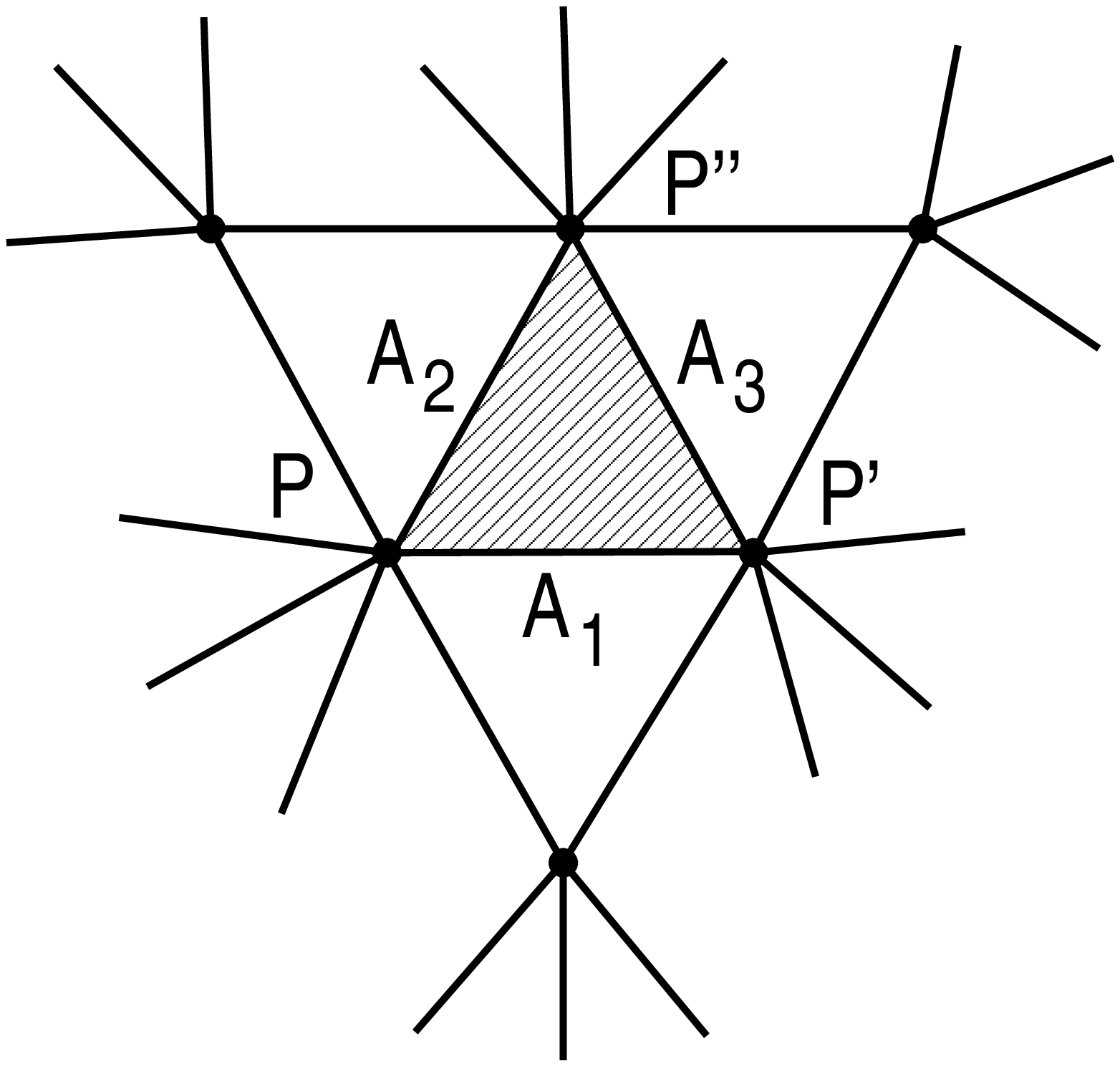}}
\end{center}
\label{fig:fig4}
\caption{}
\end{figure}

Consider now the the combined operator $Q_L=\{Q^b,Q^w\}$. It defines some discrete $GL_2$ Connection in the manifold $M^2$ associated with self adjoint real operator $L$.

\begin{thm}The Connection $Q_L$ is an $SL_2$ Connection.
\end{thm}

\vspace{0.2cm}

{\bf Proof}. Every edge $R=PP'$ is intersection of black and white triangles $R=T\bigcap T'$. From the equality $Q^{b+}Q^b=Q^{w+}Q^w$ modulo diagonal part we conclude that
every nondiagonal coefficient of operator $L$ is presented as a product twice through the coefficients of operators $Q^b$ and  $ Q^w$ correspondingly:
 $$u_{T:P}u_{T:P'}=u_{T':P}u_{T':P'}=A(PP') $$

However, it was proved above that the set of numbers $A(R)$ for every edge $R=PP'$ defines a $SL_2^{\pm}$ Connection. Every thick path on a manifold with
``discrete conformal structure'' has even number of simplices, therefore this connection is $SL_2$.
Theorem is proved.

\pagebreak

{\bf 3. Factorizations, Laplace Transformations and Discrete Analog of 2D Toda Lattice.}

\vspace{0.2cm}

Various factorization of linear operators acting on the scalar functions in $R^D$ leads to the so-called   {\bf Darboux Transformations}
where $L\psi=\lambda\psi$:
$$L=Q^+Q\rightarrow QQ^+=\tilde{L}, \psi\rightarrow \tilde{\psi}=Q\psi$$
and {\bf Laplace Transformations}--see \cite{N1,ND1,ND2,N2,N3,N4,NK, N6,N7}.

 Here $L\psi=0$
$$L=Q^+Q+W\rightarrow W^{1/2}QW^{-1}Q^+W^{1/2}+W=\tilde{L}$$
$$\psi\rightarrow \tilde{\psi}=W^{-1/2}Q\psi$$

In order to apply Laplace transformation, the product $W^{-1}Q\psi$ should be well-defined. However, the operator $ Q$ maps in several cases above the initial space of functions
 into another functional space. At the same time the potential $W$ maps the initial space into itself, so our product is not well defined in general.   The Darboux transformation
 always can be applied but we need property $W=0$ for our factorization which is not true in many cases. For $D=2$ we found following classes only where Laplace transformation
 is always well-defined and can be nontrivially iterated infinite number of times.

 {\bf I. The Square Lattice}. The operators $L$ acting in the space of functions of  vertices in the square lattice  with basic shifts $T_1,T_2$, of the ''hyperbolic form'' (see \cite{N1,ND1})
 $$L=a(m,n)+b(m+1,n)T_1+c(m,n+1)T_2+d(m+1,n+1)T_1T_2$$
 with gauge equivalence group of the ''hyperbolic equation'' $L\psi=0$:
 $$L\rightarrow fLg,\psi\rightarrow g^{-1}\psi$$
 where both functions $f,g$ are everywhere nonzero. We have 2 factorizations (''right'' and ''left'')  $$L=f[(1+uT_1)(1+vT_2)+w]=g[(1+vT_2)(1+uT_1)+w_L]$$
   Laplace transformation is defined as usual $$\tilde{L}=f'[(1+vT_2)w^{-1}(1+uT_1)+1]$$

 In order to apply Laplace Transformation, we use   representative $L$ in the zero level gauge class where $L=(1+uT_1)(1+vT_2)+w$. Here $u=u(m+1,n),v=v(m,n+1),w=w(m,n)$. For the Laplace Image $\tilde{L}$ corresponding coefficients are $u',v',w'$. They are taken in the same points.

 We have to choose $f'$ properly to return to the similar gauge form after transformation  $f'=(1+w')w/(1+w)$:
 $$\tilde{L}=f'[(1+vT_2)w^{-1}(1+uT_1)+1]=(1+u'T_1)(1+v'T_2)+w']$$
Here we use notations for the shifts of functions $u_i=T_i^*(u),v'_i=T_i^*(v'),...$.

 After substitution, we have: $$u'=f'u/w,v'=f'v/w_2,u'v'_1=f'vu_2/w_2    $$
 We still have gauge transformation preserving this form $L\rightarrow f^{-1}L f, \psi\rightarrow f\psi$.
The invariants of the last  gauge transformations are following: The potential $w$  in the factorized  form
and {\bf  Curvature} $H=vu_2u^{-1}v_1^{-1}$ similar to magnetic field in the continuous case. After Laplace Transformation we get result $$H'=(1+w'_2)/(1+w_2), 1+w'=(1+w)w_{m-1,n}w_{m,n+1}w^{-1}w_{m-1,n+1}^{-1}H_{m-1,n}$$
or after shift $T_1$:
$$1+w'_1=(1+w_1)ww_1^{-1}w_{1,2}^{-1}H$$
So, following the work \cite{ND1}\footnote{There were  mistakes in the formulas for the equation in the  work \cite{ND1} corrected here.}, we can express the invariants $H',w'$ through the invariants $H,w$ only
similar to the continuous case. We exclude $H, H'$ expressing all Chain through the variable $w$(see \cite{N2}).
 We are coming  to the discretization of the 2D Toda Lattice
  following the classical scheme of Darboux and his school in the XIX Century and based on the Laplace Chains (see quotations in the work \cite{N2}).  So  the Laplace Chain $$...\rightarrow L_{k-1}\rightarrow L_{k}\rightarrow L_{k+1}\rightarrow...$$
is described described by the  equation
$$\frac{w''_1+1}{w'_1+1}\times \frac{w_2+1}{w'_2+1}=\frac{w'w'_{1,2}}{w'_1w'_2}$$
 where $f'$ means the same function for the result of Laplace Transformation, $f''$ means the same made twice.
 In particular $w,w',w''$ correspond to the numbers $k-1,k,k+1$ in our chain $w^k(m,n)=w',w^{k-1}(m,n)=w,w^{k+1}(m,n)=w''$ in the point $(m,n)$, $w_i$ means  shift of this function to the direction $T_i, i=1,2$, i.e. $w_{m+1,n}$ and $w_{m,n+1}$ correspondingly: so the equation for the quantity $w^k(m,n)$ is following
 $$\frac{w^{k+1}_1+1}{w^{k}_1+1}\times\frac{w^{k-1}_2+1}{w^k_2+1}=\frac{w^kw^k_{12}}{w^k_1w^k_2}$$

{\bf Example}. Following \cite{ND1}, consider a Cyclic Chain Problem here for the case of period equal to 2
leading in the continuous case to the sin(sinh)-gordon equations. Let $a=w^{2k},b=w^{2k+1}$ for all $k$.
We get 2 equations for $a(m,n)$ and $b(m,n)$. They imply  that for the quantity $G=ab$ we have
$$\frac{GG_{12}}{G_1G_2}=1$$
  Consider now only its partial solution $G=C$ where $C$ is a constant. Finally we are coming to the system
  $$\frac{C+b_1}{1+b_1}\times \frac{C+b_2}{1+b_2}=bb_{12}$$
  It can be viewed as some discrete analog of the sinh-gordon system as it was pointed out already in \cite{ND1} in 1997. For $C=1$ this system degenerates to the trivial form $bb_{12}=1$.

 The equivalence of this discretization of the 2D Toda Lattice with known systems is following: its connection with totally discrete 3D  Hirota system was   was done by polish group (see below).

 The  family of Hirota systems can be written in the  form

 $$\gamma F(k+1,m+1,n)F(k-1,m,n+1)+\alpha F(k,m,n)F(k,m+1,n+1)+$$
 $$+\beta F(k,m+1,n)F(k,m,n+1)=0$$
assuming that $\alpha+\beta+\gamma=0$.

 It can be  written in the form similar to our system if $\gamma\neq 0$:
$$\frac{v^{k+1}_1v^{k-1}_2}{v^k_1v^k_2}=\lambda+(1-\lambda)\frac{v^kv^k_{1,2}}{v^k_1v^k_2}$$
where $F(k,m,n)=v^k(m,n)$ and $v^k_i=T_i^*(v^k) $ as above are  shifts  along the directions $m,n$ in the selected plane.

The discretization of  2D Toda system presented above also can be written in the similar form

$$\frac{v^{k+1}_1v^{k-1}_2}{v^k_1v^k_2}=
\frac{(v^k-1)(v^k_{1,2}-1)}{(v^k_1-1)(v^k_2-1)}$$

Here $v^k(m,n)=w^k(m,n)+1$ in the previous notations.  After scaling transformation $v\rightarrow \kappa v$ we can rewrite this system in the form $$\frac{v^{k+1}_1v^{k-1}_2}{v^{k}_1v^k_2}=\frac{(v^k-\kappa)(v^k_{1,2}-\kappa)}{{v^k_1-\kappa)}{v^k_2-\kappa)}}$$
with $\kappa\neq 0$.  In the limit $\kappa\rightarrow 0$ we are coming to the partial degenerate case of Hirota systems. Independently similar approach was developed in the work \cite{Do1}. However, few years later these authors
constructed the isomorphism of this system with Hirota System (more precisely, with the {\bf ''Three-terms Hirota System''}in the work \cite{Do2}).

 \pagebreak

 {\bf II. The Trivalent Tree}. The real selfadjoint 4th order operators $L$ acting on the functions of vertices in the trivalent tree $\Gamma$--see\cite{NK}. We have here $$ L\psi_P=\sum_{P''}a_{PP''}\psi_{P''}+\sum_{P'}b_{PP'}\psi_{P'}+W_P\psi_P$$
 where $PP'P''$ is a short path of the length 2, $|PP'|=1$ and $|P'P''|=1$--see Fig 5

 \vspace{0.3cm}

\begin{figure}
\begin{center}
\mbox{\epsfxsize=6cm \epsffile{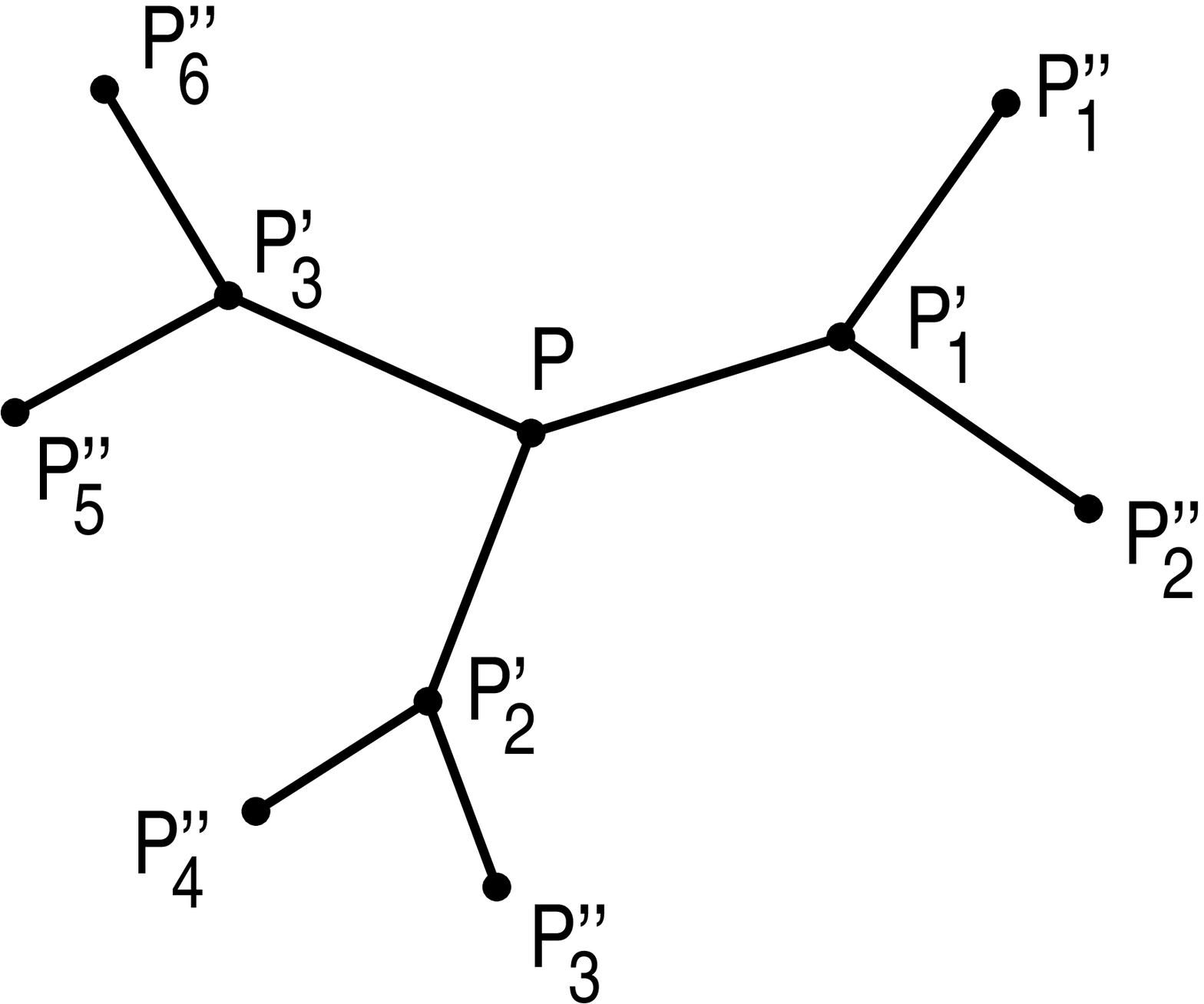}}
\end{center}
\label{fig:fig5}
\caption{}
\end{figure}

 It always can be factorized in the form $L=Q^+Q+u_P$ where $$Q\psi_P=\sum_{P'}d_{PP'}\psi_{P'}+v_P\psi_P$$
  $$a_{PP''}=d_{P'P}d_{P'P''}, b_{PP'}=d_{P'P}v_{P'}+d_{PP'}v_{P}$$
  $$W_P=v_P^2+\sum_{P'}d^2_{P'P}+u_P$$

  Laplace transformation is defined as usual $$\tilde{L}=Qu^{-1}Q^++1,\tilde{\psi}=Q\psi$$
  in the self-adjoint zero level equivalence class $L\rightarrow fLf, \psi\rightarrow f^{-1}\psi$. The iteration of Laplace transformation
  leads to  the Laplace Chains
  $$...\rightarrow L_n\rightarrow L_{n+1}\rightarrow ...$$
  where $\tilde{L}_n=L_{n+1}$.
    In this case the choice of factorizations depends on  parameter  because the equation
  $$b_{PP'}=d_{P'P}v_P+d_{PP'}v_{P'}$$ has an one-parametric family of solutions for $v$ depending on the initial value
  $v_{P_0}$ in the selected point $P_0\in \Gamma$. We perform Laplace transformations every time choosing different parameters for the solution of this equation.  Therefore Laplace Chains may be nontrivial and long. They are similar to the Darboux Chains in the 1D case of the continuous  Sturm-Liouville Operators studied by J.Weiss, A.Shabat and A.Veselov \cite{W,S,SV}, where factorization also depends on parameter numerating solutions of the Riccati Equation.
We are planning to  investigate corresponding analogs  of discrete Toda Lattice based on the sequences of the Laplace-Darboux Chains. According to our impression, these systems are much more simple than in the cases above.

{\bf  III. The Equilateral Triangle Lattice}. The real self-adjoint second order operators $L$ acting in the space of functions of vertices in the equilateral triangle lattice with shifts $T^{\pm 1}_1, T_2^{\pm 1}, (T_1^{-1}T_2)^{\pm 1}$
 of equal length $$L=a(m,n)+\{b(m+1,n)T_1+c(m,n+1)T_2+$$
 $$+d(m-1,n+1)T_1^{-1}T_2 + (adjoint)\}$$
 where $T_i^+=T_i^{-1}$
We have here right and left factorizations
$$L=Q^+Q+W,  Q=(u+vT_1+wT_2)$$ and $$L=Q'^+Q'+W', Q'=u'+v'T_1^{-1}+w'T_2^{-1}$$
This lattice defines naturally a black-white colored triangulation of the plane $R^2$ where $Q=Q^b$ and $Q'=Q^w$--see Fig 6.

\pagebreak

\begin{figure}
\begin{center}
\mbox{\epsfxsize=6cm \epsffile{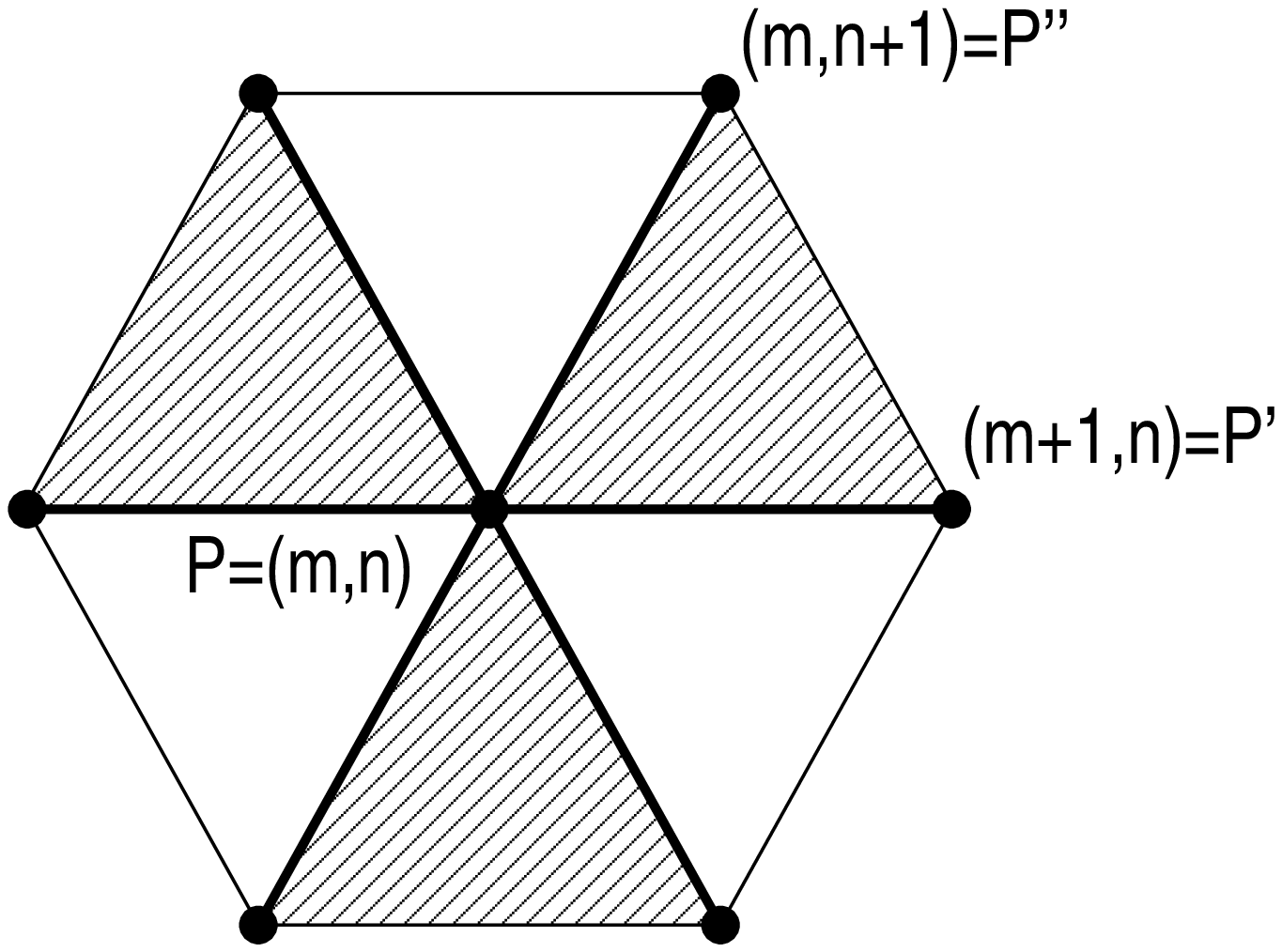}}
\end{center}
\label{fig:fig6}
\caption{}
\end{figure}

\vspace{0.1cm}

 The coefficients $b,c,d$
correspond to the edges $R=PP',PP'',P'P''$ where $P=(m,n),P'=(m+1,n),P''=(m-1,n+1)$.

This is a partial case of the triangulated 2-manifold with black-white coloring of triangles--see paragraph 1. We have here a natural isomorphism between the sets of black and white  triangles with set of vertices $(m,n)$. Therefore our operators $Q^+,Q$ and $Q'^+,Q'$ map the space of functions of vertices into itself.
Laplace transformations can be applied here infinite number of times.
  Infinite Chains of Laplace transformations
are necessary to define   discretization of the 2D Toda lattice.

No gauge transformations are allowed in this case because the transformation $L\rightarrow f^{-1}Lf, \psi\rightarrow f^{-1}\psi$ leads to  nonself-adjoint operators ,
and $L\rightarrow fLf$  destroys potentials $W,W'$ where $L=Q^+Q+W=Q'^+Q'+W'$. For the discretization of 2D Toda lattice we need to consider Laplace transformations associated with all factorizations $L=Q_j^+Q_j+W_j$
 corresponding to basic shifts $T_{j,1}, T_{j,2}, j=0,1,2,3,4,5$ where $T_{0,1},T_{0,2}=T_1,T_2$ and other bases are the rotations of this one by the angle $2j\pi/6$. The bases with numbers $j$ and $j+3(mod 6)$ are inverse to each other.

  Another  choice of representative  in the gauge class making the Laplace Transformations can be also  used for discretization of the 2D Toda Lattice here reducing the operator $L$ in the  zero level gauge class $L\psi=0$ $$L\rightarrow fLf,\psi\rightarrow f^{-1}\psi$$ to the form $W=const$ with $const=0$ or $const =1$.
 Here we have $$L=Q_j^+Q_j+const\rightarrow\tilde{L}_j= Q_jQ_j^++const=\tilde{\psi}_j$$
 $$\psi\rightarrow Q_j\psi=\tilde{\psi}_j$$
 Every time we have to make new factorization and after that  perform the ''zero level'' gauge transformation reducing potential to the constant. The whole collection of Laplace transformations in this case should be optimally organized using relations between them. Anyway, we can obtain some discrete ''Toda type'' system on the 3D Lattice in $R^3$.
 Such candidate already exists since early 1980s (see \cite{Mi}. We call it ''the 4-term Hirota-Miwa System''. We think that  some people (like the authors of \cite{Do2}  from the polish group) already established
its isomorphism with Toda-type system obtained by the Laplace Transformations from the 2D discrete Schrodinger operators. So the completely integrable(CI) discrete 2D Toda type systems following from this approach already were found before in the early 1980s during the period of intensive search of new CI systems by many groups.

 \vspace{0.3cm}

\begin{thm}The black and white operators $Q^b=Q_j=u_j+v_jT_{j1}+wT_{j2}$ and $Q^w=P_j=u'_j+v'_jT_{j1}^{-1}+w'_j T_{j2}^{-1}$ as above form together the $SL_2$ discrete connection $\{Q^w,Q^b\}$ which does not depend on $j$.
\end{thm}
{\bf Proof}. Easy to see that this connection exactly coincides with connection defined by the factorization of real self-adjoint operators in the paragraph 1 written in terms of 3 different pairs of the shift operators.

\vspace{0.3cm}

By the Theorem proved in the paragraph 2 above, this connection is always $SL_2$ because the product relations
 between the coefficients of the black and white operators immediately follows from the fact that  2 different factorizations represent the same operator $L$ as it was pointed out above. So our theorem is proved.

{\bf Our Conclusion} is that canonical $SL_2$ connection is associated with real self-adjoint operator on the equilateral
triangle lattice. It is analogous to the magnetic part of Schrodinger operators in the Quantum Mechanics.

\vspace{0.3cm}

{\Large Appendix: Discrete $SL_n$-Connections for $n>2$}

\vspace{0.2cm}

Let us introduce the following notation. Consider a $n-1$-dimensional 
face $F$ of a $n$-dimensional simplex $T$. Let 
$P_0$, $P_1$, \ldots, $P_n$, $P_n\not\in F$ be the vertices of  $T$. Denote by $A(T:F)$
the product of all connection coefficients at the face $F$
$$
A(T:F)=\prod_{0\le i<n} u_{T:P_i}.
$$

\begin{thm} A discrete connection $Q$ belongs to the $SL_n^{\pm}$ class
iff there exists a gauge transformation $Q\rightarrow fQ$ such that the identity 
\begin{equation}
\label{eq:sln}
A(T':F) = A(T'',F)=A(F).
\end{equation}
takes place for all pairs of  $n$-dimensional simplices $T'$, $T''$ intersecting by 
an $n-1$-dimensional face $F$: $F=T'\bigcap T''$, $\dim F=n-1$.
Here $f$ denotes some positive function of simplices.
\end{thm}

The proof of this Theorem will be published in our article in Russian Mathematical Surveys, v.68, No. 5 (2013).

\vspace{0.2cm}

We already worked with Canonical Connection constructing discretization of Complex Analysis for $n=2$ (see \cite{ND3,GNRoma}). For every $n\geq 2$ we   know that the curvature of this connection in any point is equal to zero (i.e. it is trivial) if and only if
 simplicial star $St(\sigma^{n-2})$ of every $n-2$-simplex contains even number of vertices. We can color $n$-simplices into the white-black colors if and only if every closed thick path consists of even number of $n$-simplices. The Holonomy Group $G$ of Canonical Connection is always part of permutation group $G\subset S_{n+1}$.

  For the construction of the discrete analog of Complex Analysis
we need the property that Canonical Connection is globally flat. In that case we have black-white coloring of $n$-simplices and $n$-dimensional family of Covariant Constants $Q\psi=0, \psi\in R^n$. So the black and white operators $Q^b,Q^w$ are well-defined here, and the connection operator is their direct sum $Q=Q^b\bigoplus Q^w$.
Covariant constant $\psi$ is defined by the set of $n+1$ real numbers $\psi_j, \sum_j\psi_j=0$, in every $n$-simplex
and its vertex $j$, unified in the whole manifold by the condition $Q\psi=0$. No theory was developed yet for $n>2$.

\vspace{0.3cm}

{\Large Appendix 2: Electric chains and Laplace transformations.}

Consider any graph $\Gamma$ (i.e. a one-dimensional simplicial complex). Let $\Gamma$ be one-dimensional
skeleton of 2-complex $K$ with following property:

{\bf  1.Every edge of graph $\Gamma$ belongs to the
         boundary of exactly one 2-dimensional triangle.

        2.Every vertex belongs to at least 3 triangles.}\\
The notion of electric chain involves the {\bf conductivities} $c(I)>0$ assigned to every edge $I$.
Here the {\bf resistance} is  $r(I)=1/c(I)$. Every {\bf voltage} function $U(P)$ of vertices generates
{\bf currents} $J$ through every oriented edge $\hat I=[P_0,P_1]$
$$
J([P_0,P_1]) = c(I)\cdot (U(P_1)-U(P_0))=(C\partial^*U)([P_0P_1]), U=U(P).
$$
The current $J$ is a one-dimensional chain on the graph. We call vertex {\bf free} if boundary
$\partial J$ does not contain this vertex (the corresponding coefficient is equal to 0). The value
of function $U(P)$ at a free vertex $P$ is:
$$
U(P) =\frac{\sum\limits_{i} U(P_i) c([P,P_i])}{ \sum\limits_{i} c([P,P_i]) },
$$
where $P_i$ \mbox{are all neighbours of } $P$. So we conclude that if all vertices are free, the Voltage function $U(P)$ satisfies
to the second order linear difference equation $LU=0$. In general, the image $LU(P)$ is a {\bf Total Current} through the vertex $P$.

{\bf The star-triangle transformation}\footnote{It is likely that this transformation was first
introduced in 1899 by Kennelly \cite{Ke} for classical electric circuits.} for one triangle $P_1$, $P_2$, $P_3$ with conductivities
$c_3$ for $[P_2,P_1]$,  $c_1$ for $[P_3,P_2]$, $c_2$ for $[P_1,P_3]$ is the following:
we put a new vertex $T$ in the center of triangle $T$ (let us call them ''black triangles'') and join it with all three vertices
$P_1$, $P_2$, $P_3$ by edges with conductivities $c'_1$ for $[P_1,T]$,  $c'_2$ for $[P_2,T]$,
$c'_3$ for $[P_3,T]$ :
$$
c'_1(T)=\frac{c_1 c_2+ c_1 c_3 + c_2 c_3}{c_1}, \  c'_2(T)=\frac{c_1 c_2+ c_1 c_3 + c_2 c_3}{c_2}, \
c'_3(T)=\frac{c_1 c_2+ c_1 c_3 + c_2 c_3}{c_3}.
$$
Remove now from the graph all three edges $[P_1,P_2]$, $[P_2,P_3]$, $[P_3,P_1]$. Define the new
voltage function in the new extended graph:\\
 $U'=U$ for all vertices except $T$ and \\ $\partial J'=\partial J$.\\
In particular it means, that the sum of currents entering new  vertex $T$ is equal to 0.

By {\bf the Star-Triangular transformation} in the whole complex $K$ we call the star-triangle transformation applied to all
2-dimensional (black) simplices $T$ of the complex $K$.

{\bf The Star-Triangle Transformation} is defined as a mapping from the linear space of voltage functions
$U(P)$ into the space of voltage functions $U'(T)$.

 Voltage function $U(P)$  with free vertices only satisfies to the following
difference equation in the original graph $\Gamma$ with vertices $P_i$:\\
$$LU=0, L= \partial C \partial^*, C:I\rightarrow c_I I$$
for all edges $I$.

Consider the {\bf Black Triangle Operator} $Q^b\psi(T)=\sum c_i'\psi(P_i), i=1,2,3$ mapping functions of vertices in the functions of black triangles as it is defined in the works of the present authors and I.Dynnikov since 1997.

{\bf Theorem 1}.The operator $L=\partial C \partial^*$ can be factorized in the black triangle (Novikov-Dynnikov) form
$$L=Q^+(C')^{-1}Q-W, \ \ C':T\rightarrow (\sum_{i=1}^{i=3} c'_i(T)) T$$ for all black triangles $T$\\
The proof can be easily checked by the direct substitution.  In the works of Novikov and Dynnikov  (see \cite{ND1,ND2}) factorization
$L=P^+P-V$ was used where $P=\left(\sqrt{(C')}\right)^{-1}Q$ with black triangle operator $P$ and $\psi'=P\psi$. In our case $\psi'=(C')^{-1}Q \psi$. Here we always use the gauge condition $L\psi=0$ for $\psi=const$. The gauge group is $L\rightarrow f(P)Lf(P),\psi(P)\rightarrow f^{-1}(P)\psi(P)$.
It allows to write down this transformation in the convenient special form.

{\bf Theorem 2.} The operator $L'$ acting on the functions of the black triangles, is obtained from $L$ by the Star-Triangle Transformation in the theory of electric chains. It has a form
$$L'= QW^{-1}Q^+-C'$$ and $LU=0$ implies $L'U'=0,U'=C'QU$. So the operator $L'$ is obtained from the operator $L$ by the Novikov-Dynnikov (ND) Laplace-type transformation (written in the special gauge form) for every complex $K$ consisting of black triangles, defined above.

\vspace{0.2cm}

This appendix appeared after the information about Star-Triangle Transformations in the classical theory of electric chains communicated by Korepanov. Korepanov and Kashaev used it to construct some solutions to the Yang-Baxter type equations--see \cite{Ka,Ko}. After discussion with Korepanov we came to conclusion that this transformation is related to the Laplace type transformation for the difference systems developed for the triangulated structures since 1990s
in our works.

\end{document}